%% file: DynamicZF-NF-Re-Refereed.tex
\documentclass[12pt]{iopart}
\usepackage{iopams}
\usepackage{color}
\usepackage{graphicx}

\newlength{\figwidth}
\newcommand{\new}{}
\newcommand{\neww}{}
\newcommand{\eqref}[1]{(\ref{#1})}

\begin{document}

\title{Dynamic transport regulation by zonal flow-like structures in the TJ-II stellarator}

\author{J. A. Alonso, C. Hidalgo, M. A. Pedrosa, B. Van Milligen and D. Carralero}
\address{Euratom-CIEMAT Assoc., Laboratorio Nacional de Fusi\'on, Av Complutense 22, 28040 Madrid, Spain}
\author{C. Silva}
\address{Euratom-IST Assoc., Instituto de Plasmas e Fus\~ao Nuclear, Instituto Superior T\'ecnico, Universidade T\'ecnica Lisboa, Lisboa, Portugal}

\ead{ja.alonso@ciemat.es}

\begin{abstract}
Floating potential structures that are correlated over a long distance are observed with a 2D probe array in the plasma edge of the TJ-II stellarator.  We introduce a method based on the Singular Value Decomposition to extract the spatio-temporal structure of the global, fluctuating, zonal-flow-like floating potential from the combined measurements of a 2D probe array and a distant single probe. The amplitude of these global structures is seen to modulate not only the high $k_\theta$ spectral power of the local turbulence, but also particle transport into the unconfined Scraped-Off Layer, as observed by $H_\alpha$ monitors around the device. 
These observations provide the first direct evidence of the global  modulation of transport by zonal flow-like structures. The ability to identify spontaneous and collective rotation events with flux surface symmetry opens up the possibility to perform unperturbative studies of the effective viscosity in stellarators and tokamaks.
%

\end{abstract}


\maketitle

\section{Introduction}
Mass flows are an active research topic  in magnetic confinement fusion due to their importance for plasma stability and confinement.
In recent years, both experimental and theoretical efforts were made to improve understanding of the momentum transport mechanisms that determine the observed plasma rotation profiles (see, e.g., \cite{TalaPPCF2007}). 
It is generally accepted that fluctuating zonal flows (ZF) in magnetically confined plasmas can be driven by turbulence through an inverse cascade process familiar from 2D hydrodynamic turbulence \cite{HasegawaPRL1987}. 
These structures are thought to regulate turbulent transport by a) draining energy from small scale turbulence, via the turbulent momentum flux (Reynolds and Maxwell stresses), and depositing it into the $k_\theta = 0$ mode (with $E_\theta = -\partial_\theta \phi = 0$ and thus $v_r = E_\theta/B = 0$ ), and b) breaking up turbulent eddies once the associated sheared flow pattern is formed. 
In this view, large-scale flows (mean and fluctuating) and small-scale turbulence form a dynamical system whose equilibrium state determines the turbulent transport levels \cite{DiamondPPCF2005}. Possibly, ZFs could explain one of the most fundamental, long-standing issues in fusion physics: the Low-to-High confinement transition (see \cite{EstradaEPL2010} for a recent experimental finding in this respect). Indeed, the practical importance of zonal flows stems from their capability to affect global cross-field transport. 

Experimentally, global, fluctuating flow patterns manifest themselves as electric potential structures exhibiting long-range correlation (LRC), and have been observed in different devices using Langmuir probes and Heavy Ion Beam probes \cite{FujisawaPRL2004, PedrosaPRL2008, XuNF2011, LiuPRL2009}. These LRC potential structures oscillate but are constant on a flux surface (which amounts to having $k_\|\approx k_\perp \approx 0$), but to qualify as a zonal flow-like structure they must also be `zonal' (i.e. have a finite radial scale $k_r\ne 0$) and be dominated by low frequencies (somewhere between the characteristic time-scales of turbulence, $\sim 10 \mu$s, and confinement, $\sim 10$ms).
The  turbulent drive of the ZF has mainly been investigated \cite{ShatsPRL2002, XuPRL2003, ManzPRL2009} using bi-spectral techniques \cite{DiamondPRL2000}, and a flow generation mechanism by three-wave coupling was identified, in accordance with rather general considerations concerning the structure of drift-wave equations \cite{DiamondPPCF2005}. 
{Although some devices have reported a reduction of the broadband  spectral power of turbulence associated with the presence of large scale flows (see \cite{SilvaPoP2008, LiuPRL2009}, \cite{FujisawaNF2009} and references therein), the crucial practical property of a ZF (i.e., its ability to regulate global transport) has not been tested experimentally  in fully developed broadband turbulence.}

In this paper, we provide empirical evidence of global transport regulation by zonal flow-like potential structures with long distance correlation. For that purpose, we present a simple method to isolate the temporal evolution of the fluctuating, collective floating potential oscillations from the combined measurements of a 2D Langmuir probe array and a toroidally separated single probe pin. 
This allows us to characterize the time-resolved dynamics and to study its effect on edge potential profiles, local turbulence, and global particle transport. We find that the  instantaneous ZF amplitude  modulates global particle fluxes dynamically, as reflected by $H_\alpha$ radiation monitors around the device. We show that the ability to identify spontaneous and collective rotation events with flux surface symmetry opens up the possibility to study effective viscosity in stellarators and tokamaks, without the need of external momentum input through NBI \cite{IdaPoP1997} or external biasing \cite{GerhardtPRL2005, PedrosaPPCF2007}.

\section{Experimental Set-up}
Experiments where carried out in the 4-period, flexible, low shear stellarator TJ-II using electron cyclotron heated plasmas. The typical magnetic field strength is $\sim 1$ T and the rotational transform $\frac{\iota}{2\pi}(a) = 1.65$ at the last closed flux surface (LCFS) (typical minor radius $a$ is 20 cm), dropping slowly to a central value of $\frac{\iota}{2\pi}(0) = 1.55$. Electron temperatures range from $1-0.8$ keV at the centre to an edge value of $100 - 50$ eV.%
%
\begin{figure}
\includegraphics[width=\figwidth]{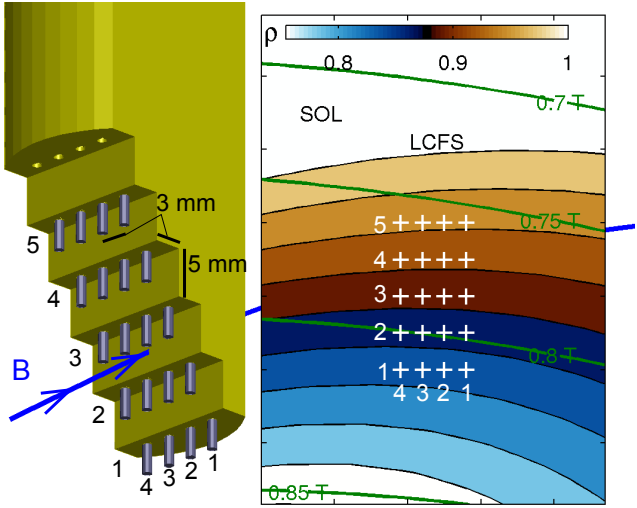}%
\caption{\label{fig:probe}Pin arrangement in the 2D probe and orientation with respect to the magnetic field structure of TJ-II.}
\end{figure}
The floating potential measurements presented in this work were obtained with a 2D array of $5\times4$ tungsten pins, with radial and poloidal separations of $6$ and $3$ mm, respectively (see figure \ref{fig:probe}). Simultaneous measurements are taken with another pin in a different sector of the machine, having a long parallel connection length with the 2D probe ($\gtrsim 100$ m).\new{The sampling rate of the floating potential signals is 2 MHz.}

\section{Spatio-temporal characterization of long range correlated structures by SVD\label{sec:svd}}
The results presented here depend crucially on the separation of the ZF contribution to the floating potential oscillations  --the long range correlated part of the potential fluctuations -- from the uncorrelated, local fluctuations. This is achieved by means of a Singular Value Decomposition (SVD), described below (an alternative method based on the same idea but using the Hilbert-Huang transform has recently been proposed \cite{CarrerasNF2011}; early applications of SVD to fluctuations in plasmas can be found in \cite{DudokPoP1994}). 

\neww{%
Given a $m\times n$ matrix $M$, its SVD consists of a product of three matrices 
\begin{equation}\label{eq:svd}
	M = U\Sigma V^{\dag}~,
\end{equation}
where $U_{m\times m}$ and $V_{n\times n}$ are unitary matrices ($UU^\dag= VV^\dag = I$) and $\Sigma_{m\times n}$ is a real, non-negative, diagonal matrix with elements $\sigma_\alpha$ sorted in decreassing order, i.e. $(\sigma_1 \ge \sigma_2\ge\ldots)$. The $U$ and $V$ column vectors,  $\{\mathbf{u}^{\alpha}\}_{\alpha=1\ldots m}$ and $\{\mathbf{v}^{\alpha}\}_{\alpha=1\ldots n}$ are orthonormal vector sets and are called \emph{topos} and \emph{chronos} respectively \cite{AubryJSP1991}. In terms of the topos and chronos vectors, equation (\ref{eq:svd}) can be written as
\begin{equation}\label{eq:svd2}
	M  = \sum_\alpha \sigma_\alpha \mathbf{u}^\alpha(\mathbf{v}^\alpha)^\dag~.
\end{equation}
}
This nomenclature is most appropriate when the matrix $M$ consists of the signals of spatially separated measurements. This is also the case here, i.e. \neww{we construct the matrix $M$ as} $M_{ik} = \phi_{fi}(t_k)$, where the $\{\phi_{fi}\}_{i=1\ldots 20}$ are the 20 floating potentials measured with the 2D probe and $\phi_{f21}$ is the floating potential of the distant probe. \neww{The $k$ index ranges over the number of samples in the floating potential signals ($6\times10^4$ in the dataset used below).}

The cross correlation matrix of the signals \new{for time lag $0$} can be expressed in terms of the topos vectors:
\begin{equation}\label{eq:xcorr}
	C_{ij} \equiv \sum_{k=1}^{n}\phi_{fi}(t_k)\phi_{fj}^*(t_k) = \sum_{\alpha=1}^{m}\sigma_\alpha^2u^\alpha_i (u^\alpha_j)^*~.
\end{equation}
This can be seen from
\begin{equation*}
	C_{ij} =\sum_{k=1}^{n}M_{ik}M_{jk}^* \Rightarrow C = MM^\dag~.
\end{equation*}
Inserting the SVD decomposition (Eq.\eqref{eq:svd}) one obtains
\begin{equation*}
	C = MM^\dag = U\Sigma V^{\dag}\left(U\Sigma V^{\dag}\right)^\dag = U\Sigma\Sigma^T U^\dag~,
\end{equation*}
which is the matrix form of equation \eqref{eq:xcorr}. The  total signal energy has the simple expression 
\[
E = \sum_{i=1}^m\sum_{k=1}^n |\phi_{fi}(t_k)|^2 = \tr C_{ik} = \sum_\alpha \sigma_\alpha^2~.
\]

With our naming convention of the floating potentials, the long range correlation between any one probe $i=\{1\ldots 20\}$ in the 2D array and the remote probe is given by $C_{i,21}$. Thus, according to Eq. \eqref{eq:xcorr}, one can write the LRC as a sum of the contributions of each of the topos $\alpha$
\begin{equation}\label{eq:lrc}
 C^{LRC}_i =\sum_\alpha \sigma_\alpha^2(u_{21}^\alpha)^*u_i^\alpha= \sum_\alpha {z}_i^\alpha~,
\end{equation}
we have defined a new quantity, the \emph{Partial Eigenmode Correlation} $\mathbf{z}^\alpha = \sigma_\alpha^2(u_{21}^\alpha)^*\mathbf{u}^\alpha$. \new{In view of the fact that this quantity gives the correlation of the distant probe with all the 20 signals in the 2D probe, this vector has 20 entries.} The relation \eqref{eq:lrc} simply expresses that for a particular topo $\alpha$ to contribute significantly to the LRC, it must carry a significant fraction of the signal energy ($\sigma^2_\alpha$) and involve the distant probe signal ($u_{21}^\alpha$).

\begin{figure*}
 \includegraphics[width=2.\figwidth]{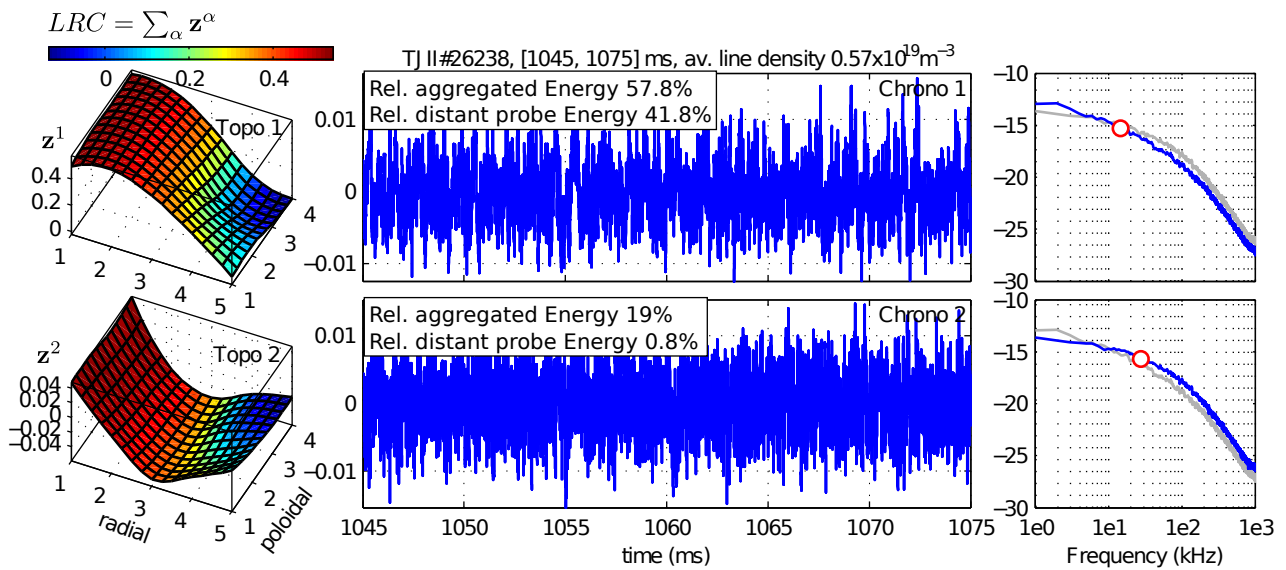}%
 \caption{\label{fig:svd}Correlation and frequency analysis of the first two SVD modes of a 30 ms record of floating potentials.  Left column: The $z$-value of the surface plot shows the Partial Eigenmode  Correlation ($z^\alpha_i$, see text) while \new{the surface coloring represents} the total long-range correlation ($C^{LRC}_i = \sum_\alpha z^\alpha_i$) \neww{and is the same in the two graphs}. Central column: chronos vectors. Right column: normalized frequency spectrum of the chronos vectors (red circles indicate the weighted mean frequency. \new{To ease their comparison the spectra of the two chronos are plotted in both graphs, the blue curve being the spectrum corresponding to the chrono on the left}).}
\end{figure*}
%
The surface plots shown in figure \ref{fig:svd} show the Partial Eigenmode  Correlation of the  first two SVD modes of a 30 ms record of floating potential measurements. The average plasma density is stable at around $\langle n_e\rangle = 0.57\times10^{19}$ m$^{-3}$, and the innermost pins of the 2D probe display a long range correlation of 0.53 (normalized to $[-1,1]$). The 20 first components of $\mathbf{z}^\alpha$ corresponding to the 2D probe array are rearranged in a square matrix, mimicking their actual spatial distribution. \neww{The surface colouring corresponds to the total LRC of the 2D pins that is given by the sum of the individual Eigenmode Partial Correlations according to Eq.(\ref{eq:lrc}).}

\begin{figure}
 \includegraphics{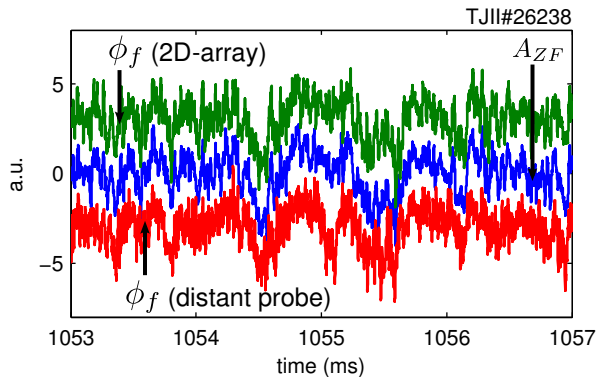}%
 \caption{\label{fig:chronos} Temporal evolution of one of the 2D probe floating potentials, the distant floating potential, and the first SVD chrono, showing an obvious mutual correlation.}
\end{figure}

This empirical decomposition reveals three characteristics features that are consistently reproduced for other similar shots having LRCs.
\begin{itemize}
\item The topo (and the LRC) displays a symmetry in the poloidal direction. While this is to be expected for sufficiently low $k_\theta$ modes it has to be stressed that this symmetry is not imposed by the SVD technique but appears naturally.
\item The first mode $\sigma^1 \mathbf{u}^1(\mathbf{v}^1)^\dag$ accounts for most of the LRC observed in the 2D array \neww{(compare the colorbar and the $z$-values of the topos, i.e. the Partial Eigenmode Correlation, in fig.\ref{fig:svd})}.
\item Ther first mode accounts for a significant part of both the cumulative signal energy $E =\sum_{\alpha=1}^{m}\sigma_\alpha^2$ (\new{namely $\sigma_{\alpha=1}^2/E =58\%$}) and the distant probe's fluctuation energy $E_{21} = \sum_{k=1}^n |\phi_{f21}(t_k)|^2 = \sum_{\alpha=1}^{m}\sigma_\alpha^2|u^\alpha_{21}|^2$ (\new{with $\sigma_{\alpha=1}^2|u^{\alpha=1}_{21}|^2/ E_{21} = 42\%$}).
\item The frequency spectrum of the first chrono is dominated by low frequencies $\lesssim 10$ kHz.
\end{itemize}
From these observations, we conclude that the coherent floating potential evolution given by the first SVD mode is a  global, low-frequency-dominated, low-$k_\theta$ electric potential structure that we will refer to as the \emph{Zonal Flow} (ZF) component. In matrix form
\begin{equation}\label{eq:zf}
	{\phi}_{ZF} = \sigma^1 \mathbf{u}^1(\mathbf{v}^1)^\dag
\end{equation}
Therefore, the first chrono will be tentatively identified with the evolution of zonal flow amplitude, symbolically $A_{ZF}(t_k) = (v^1_k)^*$.
Subtraction of the ZF component from the total 2D floating potential evolution provides a local, largely uncorrelated turbulent component given by 
\begin{equation}\label{eq:local}
\phi_\mathrm{local} = \phi_f - \phi_{ZF} 
\end{equation}
Both ZF and local components are extracted at the same time resolution as the original signals. Also, one should keep in mind that the identification made in equation \ref{eq:zf} is not absolute in any sense  (hence: zonal flow-like), but rather, motivated by the observations made above. The conditional averages and correlation analyses presented in the next section confirm this interpretation.
Figure \ref{fig:chronos} shows the ZF amplitude $A_{ZF}(t)$ (i.e. the first SVD chronos) alongside one of the 2D-array floating potentials and the distant probe signal. $A_{ZF}(t)$ clearly reflects most of the temporal evolution of either potential.

\section{Global modulations of floating potential profile and ZF dynamics}
\begin{figure}
 \includegraphics{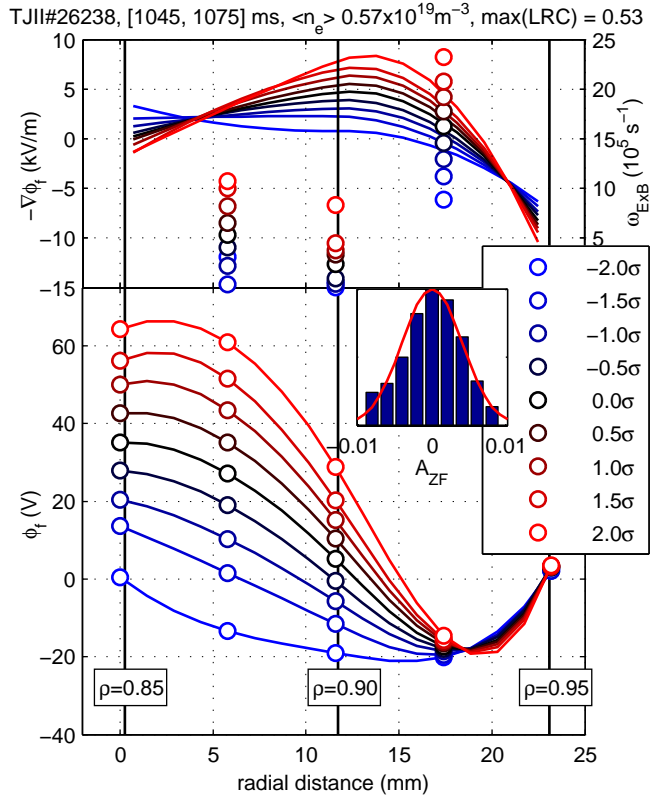}%
 \caption{\label{fig:profiles} Averaged radial profile of floating potentials measured with the 2D probe for different ranges of ZF amplitude. The inset shows the probability density function of the ZF amplitude. Instantaneous profiles falling within each of the nine bars are averaged and correspond to the entries in the legend, where $\sigma$ is the standard deviation of the ZF amplitude. The top plot shows the corresponding $E_r$ (curves, left axis) and shearing rates (circles, right axis).}
\end{figure}
The modulation of the floating potential profiles $\phi_f(r)$ by the ZF and their derived electric fields $E_r = -\frac{d\phi_f}{dr}$ and shearing rates $\omega_{E\times B} = \left|\frac{1}{B}\frac{d^2\phi_f}{dr^2}\right|$  are shown in figure \ref{fig:profiles}. Time slices are selected corresponding to specific ranges of $A_{ZF}$ values, and the floating potentials are time-averaged over those slices. \new{The inset in figure \ref{fig:profiles} shows the probability distribution of $A_{ZF}$ whose values are given in the normalised units used to represent the chronos vectors. The floating potential modulations show how this value translate into Volts at the different locations.}
Thus, we find that positive zonal flow amplitudes increase the $E\times B$ shearing rate, reaching values over $2\times10^6$ s$^{-1}$ at $\rho\sim 0.92$, whereas negative amplitudes act to flatten the mean potential profile, reducing the shearing rate.  The overall mean profile coincides with the  black curve, corresponding to the zero ZF amplitude. The maximum ZF amplitude is located around $\rho=0.87$ near the second innermost row. This is confirmed by measurements in which the probe was located further inside the plasma column. From the profile modification, a rough estimate of the ZF radial scale is $\sim 2$ cm $\approx 20\times r_{Li}$ (ion Larmor radius).
%
The oscillating radial electric field computed from the floating potential gradient varies from $0.5$ to $8$ kV/m at the location of the peak $E_r$. The $E\times B$ poloidal velocities obtained from this $E_r$ estimate range from $0.6$ to $10$ km/s.

It should be noted that these estimates of $E_r$ and $\omega_{E\times B}$ do not take into account the electron temperature contribution to the plasma potential. \neww{Besides, all but the innermost pins in the 2D probe are shadowed by the insulating Boron Nitride probe body, which may introduce further (although smaller) corrections to the mean plasma potential caused by the density wake of the parallel ion flow past the probe \cite{GunnPoP2001}. However their observed modulations or relative changes are less sensitive to these corrections and are considered a good proxy of plasma potential modifications.}

\begin{figure}
 \includegraphics{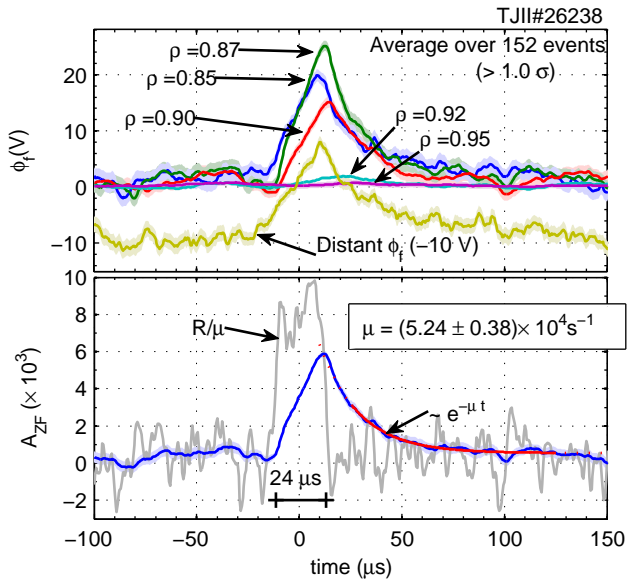}%
 \caption{\label{fig:condAzf}Top: Conditionally averaged floating potentials (the conditional trigger is $A_{ZF}/\sigma > 1$) showing a similar evolution of the 2D-array and distant probes (average values have been subtracted showing differences w.r.t. the mean potential profile -- cf. figure \ref{fig:profiles}). Bottom: Dynamical parameter of a ZF burst extracted from the conditionally averaged $A_{ZF}$ pulse shape and the model given by Equation \ref{eq:model}. \new{$A_{ZF}$ units are arbitrary. The peak amplitude ($6\times10^{-3}$) correspond to about $20$V increase in the inner floating potentials.}}
\end{figure}
These oscillations are collective for they are observed in both probes. This is illustrated by the floating potential evolution shown in figure \ref{fig:condAzf}, conditionally averaged \new{over 152 events} with $A_{ZF}/\sigma>1$. The response of the distant probe has the same waveform as that observed with the 2D array.
%
The average waveform of the zonal flow burst can be used to estimate characteristic parameters of the ZF dynamics. We model its time evolution with the simple 1D differential equation
\begin{equation}\label{eq:model}
	\frac{d}{dt}A_{ZF}(t) = R(t) - \mu A_{ZF}(t) 
\end{equation}
where $R(t)$ is the ZF drive \new{(which may depend on $A_{ZF}$)} and $\mu$ is the damping rate due to {constant} viscous forces. An estimate of $\mu$ is obtained from an exponential fit to the tail of the ZF decay in the $[30, 120] \mu$s interval (figure \ref{fig:condAzf}), assuming the flow is freely decaying ($R\approx 0$). The resulting decay time is $\tau = (20 \pm 2) \mu$s, similar to the decay time measured in earlier work \cite{PedrosaPPCF2007} using external biasing switch-off experiments. Note that this damping rate is a measure of effective viscosity corresponding to the flux surface. Using this value, we also obtain the time evolution of the forcing term $R(t)$ by integration of Eq.\eqref{eq:model}. Figure \ref{fig:condAzf} shows that the waveform of $R$ if almost square. Rise and decay times are both $\sim 3 \mu$s for a total boost duration of $24\mu$s. In physical units, the incremental force (acceleration) and asymptotic velocity are respectively $R(0)= 2 \times 10^8$ m/s$^2$, and $R(0)/\mu = 4$ km/s.

Equation \eqref{eq:model} can be interpreted to model the evolution of the flux surface averaged poloidal velocity (see, e.g., \cite{DiamondPPCF2005}), where $R$ is the turbulent drive (associated with the Reynolds stress) and the $-\mu$ term accounts for the neoclassical viscosity or magnetic pumping effect. To obtain the dynamical model parameters we had to rely on the hypothesis that the forcing term is small in the region of the fit, as suggested by the exponential waveform typical of freely decaying flows (see \cite{PedrosaPPCF2007}, \cite{GerhardtPRL2005}), and assume that the viscous coefficient is constant in time. \new{Future work will investigate both the relation between the ZF drive and the locally measured Reynolds stress, and the relation between the observed damping time and the effective neoclassical flow decay rate.}

\section{Modulation of local high-$k_\theta$ turbulent spectrum and effects on global particle fluxes}
%
\begin{figure}
 \includegraphics{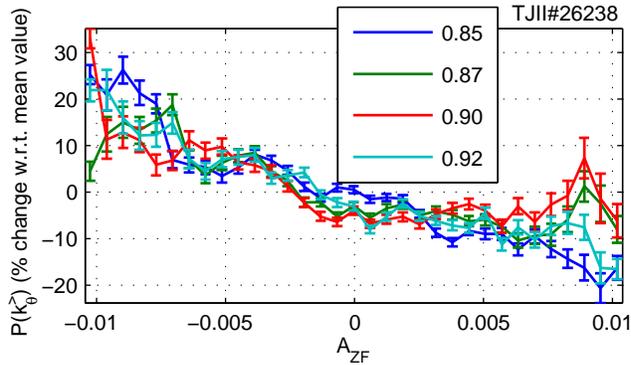}%
 \caption{\label{fig:highk} Expected values of power in the high $k_\theta$ part of the spectrum as a function of the ZF amplitude. The absolute spectral power in these wavelengths increases for decreasing radius. To simplify comparison, we have normalized the spectral power at each radius to the corresponding time average power, and expressed the result as percent difference w.r.t. this value.}
\end{figure}
The 4 poloidally separated pins of the 2D probe allow us to obtain the instantaneous spectral power in the $k_\theta = 2\pi/\lambda_\theta \approx 0, 5$ and $10$ cm$^{-1}$ wave-vectors. In figure \ref{fig:highk}, we plot the expected value of the spectral power of the uncorrelated part of the floating potential in the two non-zero $k_\theta$ wave numbers, $P_{k_\theta^>} = P_{(5\,\mathrm{cm}^{-1})} + P_{(10\,\mathrm{cm}^{-1})}$, as a function of the ZF amplitude for several radial locations. (That is, $E(P_{k_\theta^>}|A_{ZF})$ where the spatial Fourier transform is applied to the corresponding entries of the local floating potential $\phi_\mathrm{local}$ only, as defined in Eqs. \eqref{eq:zf} and \eqref{eq:local}).
It is found that positive ZF amplitudes are correlated   with a $\sim 15\%$ reduction in the small scale fluctuations, while a comparable increase is observed for negative ZF amplitudes. This is consistent with the profile modifications expected in either situation (fig.\ref{fig:profiles}), as positive ZF amplitudes increase the shearing rates, while negative ZF amplitudes decrease them.

\begin{figure}
 \includegraphics{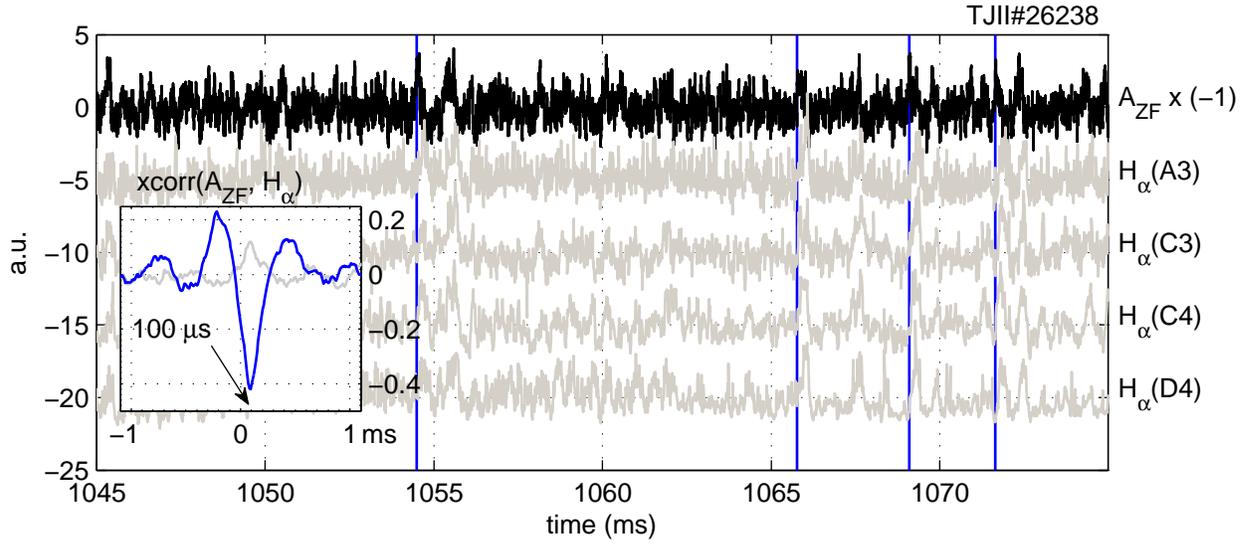}%
 \caption{\label{fig:halphas}Cross correlation curves between the zonal flow amplitude and different $H_\alpha$ monitors around the TJ-II stellarator (inset, blue line). TJ-II has four equivalent quadrants, labelled A to D, and each quadrant is subdivided into 8 sectors. The $H_\alpha$ signals are labelled according to their quadrant and sector. The cross correlation between the uncorrelated part of a floating potential (refer to equation \ref{eq:local}) and the $H_\alpha$ signals is low (inset, grey line).}
\end{figure}
The effect of these collective radial electric field modulations in the global particle transport becomes apparent in figure \ref{fig:halphas} where the ZF amplitude is plotted together with different monitors of $H_\alpha$ radiation  distributed around the device. $H_\alpha$ is mainly emitted by neutrals recycled from the walls when outward particle fluxes reach the limiters and other  material vessel components. Clearly, specific ZF relaxation events are followed by a burst of $H_\alpha$ radiation in all monitors. This prominent visual impression is confirmed by the cross correlation of the zonal flow amplitude and an $H_\alpha$ trace shown in the inset. The observed 100 $\mu$s time lag in the correlation curve implies an effective radial velocity of $100$ m/s, assuming that the plasma-wall distance is $\sim$1 cm. This velocity matches the typical radial velocity obtained from the turbulent $\tilde{E}\times B$ convective transport. On the same axis, we also plot the cross correlation of an $H_\alpha$ signal with the uncorrelated part of the floating potential measured at one of the innermost pins, showing a low positive correlation level. Overall, these results suggest that the extraction method presented above is successful in identifying ZFs that modify radial transport.

\neww{%
The results shown here focus on the detailed analysis of one shot, however similar behaviour (collective profile modulations anticorrelated with $H_\alpha$ radiation) is observed for other shots with a similar density (around $\sim 0.6\times10^{19}$ m$^{-3}$) displaying long range correlations. In previous work \cite{HidalgoEPL2009}, it was shown that LRCs in TJ-II emerge in the vicinity of this value (where the mean electric field changes from a positive to a negative root) and at high densities ($\sim 2\times10^{19}$ m$^{-3}$) close to the L-H transition. The investigation of the spatio-temporal characteristics of the correlated structures at high density is left to future work.}

%
%
%
%

\section{Conclusions}
We have introduced a general method to extract the spatio-temporal evolution of global flow structures from the time traces of several `nearby' floating potential measurements and a `distant' one. This collective mode, dominated by low frequencies (and zonal flow-like) is seen to modulate the high $k_\theta$ component of the turbulent spectrum and, for the first time, has been shown also to modulate global outward particle fluxes, as monitored by various $H_\alpha$ detectors around the TJ-II stellarator. The ability to identify spontaneous and collective rotation events with flux surface symmetry opens up the possibility to perform unperturbative studies of the effective viscosity in stellarators and tokamaks.

\section*{Acknowledgements} 
The work of J.A.A. was funded by an EFDA Fusion Researcher Fellowship (Contract Nr. FU07-CT-2007-00050).

\section*{References}

\input{DynamicZF-NF-Bib.bbl}

\end{document}

%% file: DynamicZF-NF-Bib.bbl
\providecommand{\newblock}{}